\documentclass[12pt]{article}

\begin{document}

%\hfill{Version 6: \today}

\begin{center}

\phantom{n}

\vskip 4cm

{\bf\Large Heavy-Ion Physics at the LHC}

\vskip 4cm

\end{center}
This document\footnote{On Dec.\ 20, 2000, a group of interested people 
met at LBNL to examine the possibility and prospects of US participation 
in heavy-ion experiments at the LHC. The conclusions, incorporating 
considerable input from the community at large, are presented in this 
white paper.} is to be presented at the Town Meeting at Brookhaven 
National Laboratory, Jan. 21-23, 2001, and made available to NSAC to 
aid in the long range planning process.

\newpage

%%%%%%%%%%%%%%%%%%%%%%%%%%%%%%%%%%%%%%%%%%%%%%%%%%%%%%%%%%%%%%%%%%%%%%%%%%%%%
\noindent
{\bf\large 1.\ Why should the US actively participate in the LHC 
heavy-ion physics program?}
%%%%%%%%%%%%%%%%%%%%%%%%%%%%%%%%%%%%%%%%%%%%%%%%%%%%%%%%%%%%%%%%%%%%%%%%%%%%%

\bigskip

%\begin{itemize}

\noindent
In 2006, the Large Hadron Collider (LHC) at CERN will be the highest 
energy accelerator operating on Earth. Its approved experimental program 
includes a strong heavy-ion collision component, with one dedicated 
heavy-ion experiment, ALICE, and an additional heavy-ion program in CMS. 
LHC data from heavy-ion collisions at unprecedentedly high energies will thus 
begin to complement the Relativistic Heavy Ion Collider (RHIC) scientific 
program shortly after 2006. Even a moderate US participation in the LHC 
heavy-ion experiments will have a significant impact on the LHC program. 
At the same time, this will lead to a very positive feedback on the 
understanding of RHIC physics and dynamics.

The center of mass energy for heavy-ion collisions at the LHC will 
exceed that at RHIC by a factor of about 30. This provides exciting 
opportunities for addressing unique physics issues in a completely new 
energy domain:

\begin{itemize}

\item[$\bullet$] LHC-energy heavy-ion collisions provide a unique 
opportunity to study the properties and dynamics of QCD in the classical
regime. The density of the low $x$ virtual gluons in the initial state 
will be high enough for saturation to set in so that their subsequent 
time evolution is governed by classical chromodynamics.

\item[$\bullet$] Due to the higher incident energy compared to RHIC, 
semihard and hard processes will be a dominant feature at the LHC
and gross properties of the collision can be reliably calculated
using perturbative QCD. 

\item[$\bullet$] Very hard strongly interacting probes, whose attenuation 
can be used to study the early classical chromodynamic and thermalization 
stages of the collision, are produced at sufficiently high rates for detailed
measurements. 

\item[$\bullet$] 
Weakly interacting probes, such as direct photons, $W^\pm$ and $Z^0$ 
bosons produced in hard processes, will provide information about 
nuclear parton distributions at very high $Q^2$. The impact parameter 
dependence of their production is sensitive to the spatial dependence 
of shadowing and saturation effects.

\item[$\bullet$] Compared to RHIC, the ratio of the lifetime of the 
quark-gluon plasma state to the time for thermalization is expected to 
be larger by an order of magnitude so that parton dynamics will dominate
the fireball expansion and the collective features of the hadronic final 
state. 
%Large production yields of final state particles will allow for 
%precise event-by-event and correlation analyses.

\end{itemize}

\noindent
A complete picture of heavy-ion collision dynamics at high energies
requires the analysis of the complementary information gained at both 
RHIC and the LHC. US participation in both programs is essential for 
securing a stable place at the frontier of heavy-ion research for 
our scientific community. 

\newpage

%%%%%%%%%%%%%%%%%%%%%%%%%%%%%%%%%%%%%%%%%%%%%%%%%%%%%%%%%%%%%%%%%%%%%%%%%%%%%
\noindent
{\bf\large 2.\ New opportunities for heavy-ion physics at the LHC}
%%%%%%%%%%%%%%%%%%%%%%%%%%%%%%%%%%%%%%%%%%%%%%%%%%%%%%%%%%%%%%%%%%%%%%%%%%%%%

\bigskip

%%%%%%%%%%%%%%%%%%%%%%%%%%%%%%%%%%%%%%%%%%%%%%%%%%%%%%%%%%%%%%%%%%%%%%%%%%%%%
\noindent
{\bf 2.1\ General conditions of heavy-ion collisions at the LHC}
%%%%%%%%%%%%%%%%%%%%%%%%%%%%%%%%%%%%%%%%%%%%%%%%%%%%%%%%%%%%%%%%%%%%%%%%%%%%%

\medskip

The center of mass energy for collisions of the heaviest ions at the LHC 
will exceed that available at RHIC by a factor of about 30. This opens up 
a new physics domain with exciting opportunities. Historical experience 
suggests that such a large jump in available energy usually leads to new 
discoveries. Whereas the SPS has produced excited strongly interacting 
matter near the conditions for quark deconfinement \cite{QM9}, the goal 
of the heavy-ion experimental programs at RHIC and the LHC is to 
investigate the quark-gluon plasma in full detail. RHIC is the first 
machine that allows deep penetration into this new phase, creating 
quark-gluon plasmas which are sufficiently long-lived to make them 
accessible to a variety of specific experimental probes. A 
comprehensive experimental program addressing this exciting physics, 
which foresees a broad range of systematic studies and future upgrades 
to both the machine and the detectors, has been put into place and will 
dominate the US effort in this field for the coming decade. 

Heavy-ion collisions at the even higher LHC energy, on the other hand, 
will explore regions of energy and particle density which are 
significantly beyond those reachable at RHIC. At LHC energies, the bulk 
of particle production will be due to collisions between low-$x$ virtual 
gluons in the incoming nuclei. The colliding nuclei before impact may be 
described as densely packed ``gluon walls'' approaching each other at the 
speed of light. The phase-space density of low momentum gluons is 
saturated and gluon merging becomes important \cite{MQ86,McL,Mue}. 
As a result the gluon density is sharply reduced compared to low-density 
perturbative expectations, limiting in a self-consistent way the growth of 
the production cross section for low-$p_T$ minijets. This qualitatively new 
regime of ``gluon saturation'' at low $x$ and large $A$ becomes accessible 
for the first time with heavy ions in the LHC. In Pb+Pb collisions at the 
LHC, the $p_T$-scale at which gluon saturation sets in is estimated to 
be around 2 GeV \cite{Esk}. At this scale, perturbative QCD is likely to be 
applicable. For the first time the bulk of transverse energy and 
particle production as well as the initial conditions for the subsequent 
expansion of the hot matter formed in the reaction zone can thus be 
reliably calculated within perturbative QCD.

While the concept of a dense wall of virtual gluons in the incoming 
nuclei can be probed in $eA$ collisions \cite{Ian}, in $AA$ collisions 
it plays a decisive role in our thinking about the production of dense 
strongly interacting matter and its subsequent evolution into a thermalized 
quark-gluon plasma. When two nuclei collide, the phase coherence among 
their virtual partonic constituents is broken and these partons come on 
shell. Due to the large number of gluons in any given low-$p_T$ quantum 
state this process may be described by classical chromodynamics. 
Theoretical methods based on such classical concepts 
(Weizs\"acker-Williams fields \cite{McL}, classical Yang-Mills 
dynamics \cite{Muller}, etc.) have been developed in recent years and can 
be experimentally tested at the LHC. Classical Yang-Mills fields are 
known to exhibit chaotic evolution, and this may be an important mechanism 
for early thermalization of the particles produced during the initial 
stage of the collision \cite{Muller}.

The energy density of the thermalized matter created at the LHC is 
estimated to be 20 times higher than can be reached at RHIC,
implying an initial temperature $T_0$ which is greater by more than a 
factor of 2 \cite{EKa}. Due to the higher initial parton density,
thermalization also happens more rapidly, and the ratio of the 
quark-gluon plasma lifetime (i.e. the time until the first hadrons 
begin to form) to the thermalization time accordingly increases by a 
factor 10. As a result, the fireballs created in heavy-ion collisions 
at the LHC spend nearly their entire lifetime in a purely partonic 
state, widening the time window available for experimentally probing 
the quark-gluon plasma state. 

Studies of the quark-gluon plasma can be done efficiently by using 
``hard probes'' \cite{XW1}, e.g.\ high-$p_T$ jets or photons, heavy 
quarkonia, and $W^\pm$ or $Z^0$ mesons. The ``hard scale'' 
characterizing these probes is the squared momentum transfer, $Q^2$, 
necessary for their production. At the high collision energies provided 
by the LHC, the cross sections for very high $Q^2 > $\,(50\,GeV)$^2$ 
processes are large enough to comfortably allow detailed experimental 
studies. The hard probes are created at very short times, 
$\tau_{\rm hard} \sim 1/Q \leq 0.01$\,fm/$c$, and their production can 
be calculated perturbatively once the nuclear structure functions are known.
%from accompanying $eA$ or $pA$ studies.
This time is early enough to probe the classical chromodynamic stage of 
the collision during which the ``gluon walls'' decay. The bulk of the 
secondary matter materializes somewhat later, at 
$\tau_0 \sim 1/T_0 \sim 0.2$\,fm/$c$. The hard probes are embedded in 
this secondary matter and explore its properties by scattering off it 
on their way out. The time available for this ``probing'' is of the 
order of $5-10$\,fm/$c$, given by the lifetime of the hot QGP state 
or the time needed for the probe to travel through the reaction zone, 
whichever is shorter. The medium dependence of these scattering processes 
can be calculated theoretically \cite{BDMPS,Mik,Wiedemann} so that the 
properties of the medium can be inferred from the measured final state 
of the hard probe.

\bigskip

%%%%%%%%%%%%%%%%%%%%%%%%%%%%%%%%%%%%%%%%%%%%%%%%%%%%%%%%%%%%%%%%%%%%%%%%%%%%%
\noindent
{\bf 2.2\ Unique physics questions and specific probes at the LHC}
%%%%%%%%%%%%%%%%%%%%%%%%%%%%%%%%%%%%%%%%%%%%%%%%%%%%%%%%%%%%%%%%%%%%%%%%%%%%%

\bigskip

The high energy scale of the LHC makes heavy-ion collisions sensitive 
to the nuclear parton distributions, $f_A(x,Q^2)$, at very low $x$, 
$x\sim 10^{-3}$. High parton densities leading to partonic overlap, 
gluon recombination and saturation are generally expected for 
sufficiently small values of $x$ and/or $Q^2$ and large mass numbers 
$A$. In Pb+Pb collisions at the LHC, gluon saturation strongly affects 
the production of secondary partons with $p_T<2$\,GeV. Different probes
can be used to explore the consequences of gluon recombination and 
saturation on the nuclear structure functions, i.e.\ on the distribution 
of virtual quarks, antiquarks and gluons in the {\em pre-collision} wave 
function of the incoming nuclei, and to study their effect on the density 
and further evolution of the {\em post-collision} environment.

In order to probe the {\em pre-collision} state of the nuclei and to measure 
the initial nuclear wave function, one can study the production of 
direct photons, $W^\pm$ and $Z^0$ vector bosons (observed through their
leptonic decay channels), and the yields of open charm and 
beauty mesons. Direct photons are sensitive to the quark and gluon 
distributions whereas the vector bosons probe mostly the quark and 
anti-quark distributions. Open charm and beauty production measures
properties of the gluon distribution function. $W^\pm$ and $Z^0$ 
production can be studied in $pp$ collisions at RHIC, but in $AA$ 
collisions the RHIC energy is too low for an accurate measurement. 
A high-statistics study of $W^\pm$ and $Z^0$ vector bosons in $AA$ 
collisions would thus be unique to the LHC \cite{VogZ}. 

The probes mentioned previously can be used to measure shadowing, i.e.\ the 
modification of the free nucleon parton distributions by the nuclear 
medium. In addition to the unmodified structure functions, accessible 
in $ep$ and $pp$ collisions, and the homogeneous shadowing, which can 
be studied in $eA$, $pA$ and $AA$ collisions, $AA$ collisions also 
allow the investigation of the spatial dependence of shadowing in the 
plane transverse to the collision axis by studying hard probe production 
as a function of impact parameter \cite{Eme}. 

At the LHC the nuclear parton distributions will be probed at much 
lower $x$ and higher $Q^2$ than at RHIC. This added reach is critical 
for exploring the kinematic region where the partons overlap and where
the interesting saturation effects can be observed.

The {\em post-collision} environment can be studied with hard probes 
which undergo strong final state interactions. Such probes are 
high-$p_T$ quark and gluon jets, the hadronic decays of $W^\pm$ and 
$Z^0$ bosons, and heavy quarkonia. Due to their hard production 
scales, they materialize very early after the collision and are thus
embedded into and propagate through the dense environment of
softer secondary particles as it forms and evolves. Through their
interactions with these particles they measure the properties of
the evolving medium and are sensitive to the formation of a 
quark-gluon plasma. Large transverse momentum probes are easily 
isolated experimentally from the background of soft particles 
produced in the collision. The high $p_T$ of the probes ensures
that the medium effects are perturbatively calculable which 
strengthens their usefulness as quantitative diagnostic tools. 
At the LHC the production rates for jets with $p_T>50$\,GeV are several
orders of magnitude larger than at RHIC, allowing for systematic
studies with high statistics in clean kinematic regions, far 
beyond where the RHIC experiments reach their limits.

Quark jets of known energy can be produced in reactions such as 
$g + q \rightarrow q +\gamma$ or $g + q \rightarrow q + Z^0$. 
At the LHC, 
$Z^0+$jet final states become measurable and may provide a more distinctive 
signature because the $Z^0$ is free from the high background of hadronic 
decays contributing to the direct photon spectrum \cite{CMS}.
The initial energy of the quark jet can be inferred from the measured
momentum of the $Z^0$ or photon. Any energy loss of the jet suffered
by interactions with the hot medium can thus be cleanly identified \cite{Wang}.
The measured energy loss yields the opacity of the medium, i.e. the product
of the cross section between the hard probe and the partons in the medium 
and their density \cite{Mik,Wiedemann}.
In kinematic regions where the rescattering cross section can be 
reliably calculated, the opacity provides access to the parton density after 
the collision and allows one to determine how it is affected by gluon 
saturation. Similar information can be extracted from a measurement 
of the ratio of monojet to dijet final states \cite{Gyu,Lok}. 
At the LHC, both types of measurements will be observable over a wide 
kinematic range, thus allowing studies of the energy loss as a function 
of jet $p_T$.

Heavy quarkonia also provide sensitive probes of the parton 
densities after the collision. In a dense medium, the formation of
charmonium and bottonium states is inhibited since the medium screens the 
interaction between the heavy quark and antiquark. Different resonances 
``melt'' at different temperatures, such that they act as a
``thermometer'' to probe the temperature attained in the collision. 
Systematic measurement of many different quarkonium states, including 
the dependence of the suppression effect on their transverse momenta, 
are required to fully understand the system \cite{Vog,Gun}.
At RHIC, investigation of the $\Upsilon$ system will be severely limited 
by statistics while at the LHC both $c\overline c$ and $b\overline b$ 
resonances can be studied in detail. It has been argued that the
interpretation of $\Upsilon$ suppression data is cleaner than for
the $J/\psi$ due to reduced rescattering effects with hadrons during
the late collision stages, a further point in favor of studying heavy 
quarkonia at the LHC where the $\Upsilon$ family is comfortably accessible. 
Finally, we note that at the LHC secondary $J/\psi$ production from 
$B\rightarrow J/\psi+X$ becomes relevant. Since $B$ mesons are sensitive 
to the energy loss of $b$ quarks \cite{LinV}, these secondary $J/\psi$ 
mesons can be used as a further probe of jet quenching \cite{Loh}.

Higher production cross sections at the LHC also favorably affect 
the study of physics issues which are accessible in peripheral
nuclear collisions where the two nuclei do not overlap. Multiple
vector meson production will be greatly increased \cite{Spe}.
Photoproduction of top quarks in peripheral collisions becomes possible
and probes the nuclear gluon distribution function at high $Q^2$ \cite{top}. 
Finally, photoproduction of $c\overline c$ and $b\overline b$ will 
provide an alternate method for measuring the low-$x$ gluon distribution
\cite{Kle}.

We have stressed those aspects of heavy-ion physics at the LHC which 
are unique or qualitatively different from RHIC. In addition to these, 
experiments at the LHC will also study soft probes, building on what 
has been done at RHIC. The increased energy will extend measurements 
of excitation functions far beyond RHIC data, thus providing a 
significant additional lever arm. The increased final muliplicity 
densities will allow for improved measurements of many soft physics 
observables which require high statistics, especially event-by-event 
fluctuations and correlation measurements.

\bigskip

%%%%%%%%%%%%%%%%%%%%%%%%%%%%%%%%%%%%%%%%%%%%%%%%%%%%%%%%%%%%%%%%%%%%%%%%%%%%%
\noindent
{\bf\large 3.\ General Perspective. Size and scope of the proposed effort.}
%%%%%%%%%%%%%%%%%%%%%%%%%%%%%%%%%%%%%%%%%%%%%%%%%%%%%%%%%%%%%%%%%%%%%%%%%%%%%

\medskip

The LHC is the latest in a long line of particle accelerators exploring 
physics at the high-energy frontier. It is a global project with 
contributions from five continents. Its colliding proton and heavy-ion
beams will deliver the highest energies worldwide for years to come.
Furthermore, it is unclear when or if there will be another even more 
powerful hadron collider after the LHC. The US high energy physics 
community has participated in all major high energy projects and 
is presently investing 440 million US dollars into the construction of 
the LHC and its particle physics experiments, thereby securing its place
at the front of high energy physics research. Similarly, to study 
relativistic heavy-ion collisions at the frontier, the US heavy-ion 
community has always made use of or constructed the latest and best 
facilities available. The LHC should not become the first exception.

Four large experiments, covering a rich and very exciting physics program, 
have been planned for the LHC. This program is broadly accepted by the 
scientific community at large and has been approved by all relevant CERN 
Committees. Detector construction is well on its way and expected to be 
largely complete in 2005. One of the LHC detectors, ALICE, is a dedicated
heavy-ion experiment. Another experiment, CMS, has incorporated heavy-ion 
studies as an integral part of their scientific program. New data from 
heavy-ion collisions at unprecedentedly high energies will thus emerge 
shortly after the LHC turn-on in 2006, complementing the scientific 
program at RHIC. It is very important that the US be involved in {\em both} 
programs, even while RHIC, our ``in house'' facility, continues to
be at the focus of the US heavy-ion effort. In order to fully understand 
the dynamics of nuclear matter under extreme conditions the entire available 
energy range must be spanned with a study of the phenomena in both 
kinematic regimes.

Due to the heavy US commitment towards RHIC, both in terms of manpower 
and funding resources, there can be only a moderate US involvement in the 
planned heavy-ion program at the LHC. However, such involvement 
would ensure the diversity
that has always been a very strong aspect of our field. Alhough the 
LHC detector 
designs are already well advanced, a number of opportunities remain 
for a significant impact on the physics by the participation of the US 
heavy-ion community. At a recent workshop held in Berkeley on Dec.20, 2000, 
entitled ``LHC-USA'', dedicated to LHC relativistic heavy-ion physics 
and the anticipated US participation in the program, these opportunities 
were examined. It was concluded that a meaningful US involvement should be 
at the level of about 10$\%$ of the US manpower presently involved in RHIC, 
and about 10 million US dollars in financial resources. This seems to be 
a moderate price to pay for access to this totally new physics domain, 
thereby securing for the US a stable place among the leading nations in 
relativistic heavy-ion physics.

\newpage

\centerline{\bf Agenda of the ``LHC-USA'' Workshop at LBL, Dec. 20, 2000:}

\vskip 0.8cm

\begin{tabular}{ll}
\phantom{Grazyna Odyniec} & \phantom{LBL}\\
 9:00 -  9:10  Grazyna Odyniec: Welcome, organizational remarks\\
 9:10 -  9:50  Miklos Gyulassy: ``Strong (classical) and Weak (pQCD) 
Gluon Fields at LHC''\\
\vspace{0.3cm}
 9:50 - 10:30 Ulrich Heinz: ``ALICE Physics''\\
\vspace{0.3cm}
10:30 - 11:00 Coffee\\
11:00 - 11:30 Spencer Klein: ``Quarkonia at LHC''\\
11:30 - 12:10 Terry Awes: ``Direct Photons in PHOS Detector''\\
\vspace{0.3cm}
12:10 - 12:30 Tom Humanic: ``Present Ohio State Univ. Involvement in ALICE''\\
\vspace{0.3cm}
12:30 -  1:30 Lunch\\
 1:30 -  2:00 Bj\o rn Nilsen: ``Cosmic Ray Measurements that Can be Done 
              only Using ALICE''\\
 2:00 -  2:15 Larry Pinsky: ``Houston's Potential Responsibilities within 
              ALICE''\\
 2:15 -  2:25 Jo Schambach: ``On L3 Trigger for ALICE - Austin Involvement''\\
 2:25 -  2:40 Bolek Wyslouch: ``Alice TRD in Phobos''\\
\vspace{0.3cm} 
 2:40 -  3:10 Daniel Ferenc: ``Selective Observables of Direct Photons 
              Using Correlations''\\
\vspace{0.3cm} 
3:10 -  3:30 coffee\\
\vspace{0.3cm}
3:30 -  7:00 discussion of content of the document, distribution of work, 
             time scale ...\\
\vspace{0.3cm}
 7:30 dinner 
\end{tabular}

\vskip 1cm 

\centerline{\bf Contributors and Workshop participants:}

\vskip 0.8cm

\begin{tabular}{ll}
\phantom{Grazyna Odyniec} & \phantom{LBL}\\
T. Awes & Oak Ridge National Laboratory\\
H. Crawford & Lawrence Berkeley National Laboratory\\
J. Engelage & University of California at Berkeley\\
D. Ferenc & University of California at Davis\\
M. Gyulassy & Columbia University\\
D. Hardtke & Lawrence Berkeley National Laboratory\\
U. Heinz & Ohio State University\\
G. Hoffmann & University of Texas, Austin\\
T. Humanic & Ohio State University\\
P. Huovinen & Lawrence Berkeley National Laboratory\\
P. Jacobs & Lawrence Berkeley National Laboratory\\
E. Judd & University of California at Berkeley
\end{tabular}

\vskip 0.8cm

\begin{tabular}{ll}
\phantom{Grazyna Odyniec} & \phantom{LBL}\\
S. Klein & Lawrence Berkeley National Laboratory\\
V. Koch & Lawrence Berkeley National Laboratory\\
B.S. Nilsen & Ohio State University\\
G. Odyniec & Lawrence Berkeley National Laboratory\\
L. Pinsky & University of Houston \\
H.G. Ritter & Lawrence Berkeley National Laboratory\\
J. Schambach & University of Texas, Austin\\
L. Schroeder & Lawrence Berkeley National Laboratory\\
J. Symons & Lawrence Berkeley National Laboratory\\ 
R. Snellings & Lawrence Berkeley National Laboratory\\ 
R. Vogt & Lawrence Berkeley National Laboratory and UCD\\
B. Wyslouch & Massachusetts Institute of Technology
\end{tabular}

\vskip 1.5cm

\centerline{\bf Editors:}

\vskip 0.6cm

\begin{tabular}{ll}
\phantom{Grazyna Odyniec} & \phantom{LBL}\\
Terry Awes & ORNL \\
Ulrich Heinz & OSU \\
Tom Humanic & OSU \\
Spencer Klein & LBNL \\
Bj\o rn Nilsen & OSU \\
Grazyna Odyniec & LBNL \\
Ramona Vogt & LBNL and UCD 
\end{tabular}

\vskip 1.5cm

\centerline{\bf Contacts:}

\vskip 0.6cm

\begin{tabular}{ll}
\phantom{Grazyna Odyniec} & \phantom{LBL}\\
Tom Humanic & Humanic\verb|@|mps.ohio-state.edu \\
Grazyna Odyniec & G$\_$Odyniec\verb|@|lbl.gov \\
\end{tabular}

\end{document}